\documentclass[letterpaper, 10 pt, conference]{ieeeconf}  

\IEEEoverridecommandlockouts                              

\usepackage{graphics} 
\usepackage{epsfig} 
\usepackage{mathptmx} 
\usepackage{times} 
\usepackage{amsmath} 
\usepackage{amssymb}  
\usepackage{subcaption}
\usepackage{booktabs}
\usepackage{multirow}

\title{\LARGE \bf
A Deep Learning-based \textit{in silico} Framework for Optimization on Retinal Prosthetic Stimulation
}

\author{Yuli Wu$^{1}$, Ivan Kareti\'{c}$^{1,2}$, Johannes Stegmaier$^{1}$, Peter Walter$^{3}$, Dorit Merhof$^{4}$
\thanks{*This work was funded by the Deutsche Forschungsgemeinschaft (DFG, German Research Foundation) – grant 424556709/GRK2610.}
\thanks{$^{1}$Institute of Imaging and Computer Vision, RWTH Aachen University, Germany. E-mail: {\tt\small yuli.wu@lfb.rwth-aachen.de}}%
\thanks{$^{2}$NeuroTX Aachen e.V., Germany.}%
\thanks{$^{3}$Department of Ophthalmology, RWTH Aachen University, Germany.}%
\thanks{$^{4}$Institute of Image Analysis and Computer Vision, University of Regensburg, Germany.}%
}

\begin{document}

\maketitle
\thispagestyle{empty}
\pagestyle{empty}

\begin{abstract}

We propose a neural network-based framework to optimize the perceptions simulated by the \textit{in silico} retinal implant model \textit{pulse2percept}. The overall pipeline consists of a trainable encoder, a pre-trained retinal implant model and a pre-trained evaluator. The encoder is a U-Net, which takes the original image and outputs the stimulus. The pre-trained retinal implant model is also a U-Net, which is trained to mimic the biomimetic perceptual model implemented in \textit{pulse2percept}. The evaluator is a shallow VGG classifier, which is trained with original images. Based on 10,000 test images from the MNIST dataset, we show that the convolutional neural network-based encoder performs significantly better than the trivial downsampling approach, yielding a boost in the weighted F1-Score by 36.17\% in the pre-trained classifier with 6$\times$10 electrodes. With this fully neural network-based encoder, the quality of the downstream perceptions can be fine-tuned using gradient descent in an end-to-end fashion.

\end{abstract}

\section{INTRODUCTION}

Retinal prostheses aim to provide electrical stimulation to different layers of the human retina \cite{roessler2009implantation}\cite{luo2016argus}\cite{stingl2015subretinal}\cite{palanker2020photovoltaic}. Among these, the Argus{\textregistered} II retinal prosthesis system is the retinal prosthesis that is implanted most frequently \cite{luo2016argus}. Based on the psychophysical data from the patients with this epiretinal implant, Beyeler et al. implement an open source Python library \textit{pulse2percept}~\cite{michael_beyeler-proc-scipy-2017} with different computational models, such as \cite{beyeler2019model}\cite{granley2021computational}. In this work, we rely on the \textit{Axon Map}~\cite{beyeler2019model} computational model as the hypothetical ground-truth and propose a fully neural network-based \textit{in silico} framework to generate better percepts.

Some approaches seek to optimize the electrical stimulation for the retinal implant with non-deep learning methods. 
Shah et al.~\cite{shah2019optimization} propose a closed-loop linear reconstruction framework using a greedy algorithm to optimize the electrode stimulation with an efficient number of electrodes.
More recently, Fauvel and Chalk~\cite{fauvel2022human} apply a preference-based binary Bayesian optimization algorithm \cite{brochu2010bayesian}\cite{fauvel2021efficient} to fine-tune the patient-individual perceptual model with feedbacks from the human subjects in the visual tasks. 
They use the open-source Python library \textit{pulse2percept} by Beyeler et al.~\cite{michael_beyeler-proc-scipy-2017} to simulate the functionality of the retinal implant and also to limit the search space of the perceptual model parameters.
During the processing, an optimal stimulus-wise linear encoder is found. 

As the first (in terms of the regulatory approval both in the EU and in the USA) retinal prosthesis, the Argus{\textregistered} II retinal prosthesis system emerged prior to the ubiquity of deep learning techniques.
In the era of deep learning, convolutional neural networks (CNN) became the default option for many tasks~\cite{lecun2015deep}, including those related to visual prostheses~\cite{hu2021explainable}\cite{beyeler2022towards}\cite{van2022biologically}. 
Accordingly, we introduce below some approaches working on the optimization of the retinal
prosthetic stimulation using deep learning techniques.
An encoder-decoder neural network is proposed by van~Steveninck~el~al.~\cite{van2022end}, mapping the stimulus from the input image to the simulated phosphene vision (SPV) representation through a scoreboard-based phosphene simulator. 
Additionally, Relic et al.~\cite{relic2022deep} propose a similar framework with an \textit{Axon Map} implant model (or phosphene model) based on the data of the real implant users \cite{beyeler2019model}\cite{erickson2021blind}, which has been approximated by a single-layer feedforward neural network (FNN).
Granley et al.~\cite{granley2022hybrid} introduce a hybrid neural autoencoder (HNA) to generalize the end-to-end stimulus encoding to any sensory neuroprosthesis with a novel additive reconstruction loss function, combining a mean absolute error term, a VGG similarity term~\cite{simonyan2014very}, and a Laplacian smoothing regularization term.

In this paper, we propose an end-to-end framework to optimize the phosphene vision simulated by the \textit{in silico} retinal implant model \textit{pulse2percept} using deep learning techniques. The overall pipeline consists of a trainable CNN encoder, a feedforward CNN retinal implant model and a feedforward VGG classifier. 
 The performance of the trainable CNN encoder is evaluated with a \textit{recognition} task (Section~\ref{sec:loss}).
Despite some similarities, there are many differences between our approach and those of \cite{van2022end}\cite{relic2022deep}\cite{granley2022hybrid}. 
First, we use a classifier to quantitatively evaluate the performance of the encoder with a cross-entropy loss instead of with a reconstruction loss (usually pixelwise mean squared error) as in \cite{van2022end}\cite{relic2022deep}\cite{granley2022hybrid}. 
Second, the resolutions at the interfaces are constrained. The resolution of the retinal implant is chosen as low as the Argus{\textregistered} II device (6$\times$10), which is smaller compared to 15$\times$15 as in \cite{granley2022hybrid} or 32$\times$32 as in  \cite{van2022end}. The bottleneck at the predicted percept is kept as 28$\times$28, which is narrower than 49$\times$49 as in \cite{granley2022hybrid} or 256$\times$256 as in \cite{van2022end}. 
Third, the neural networks are deliberately selected with different complexities. The U-Net~\cite{ronneberger2015u} is more efficient and suitable for an image-to-image task than the fully-connected network in \cite{granley2022hybrid}, and more powerful than the shallower CNN in \cite{relic2022deep}. The shallow VGG-5 classifier \cite{simonyan2014very} serves as a dummy brain, which evaluates the performance from the upstream without empowering generalizability.

\begin{figure*}[t]
    \centering
  \includegraphics[width=0.97\textwidth]{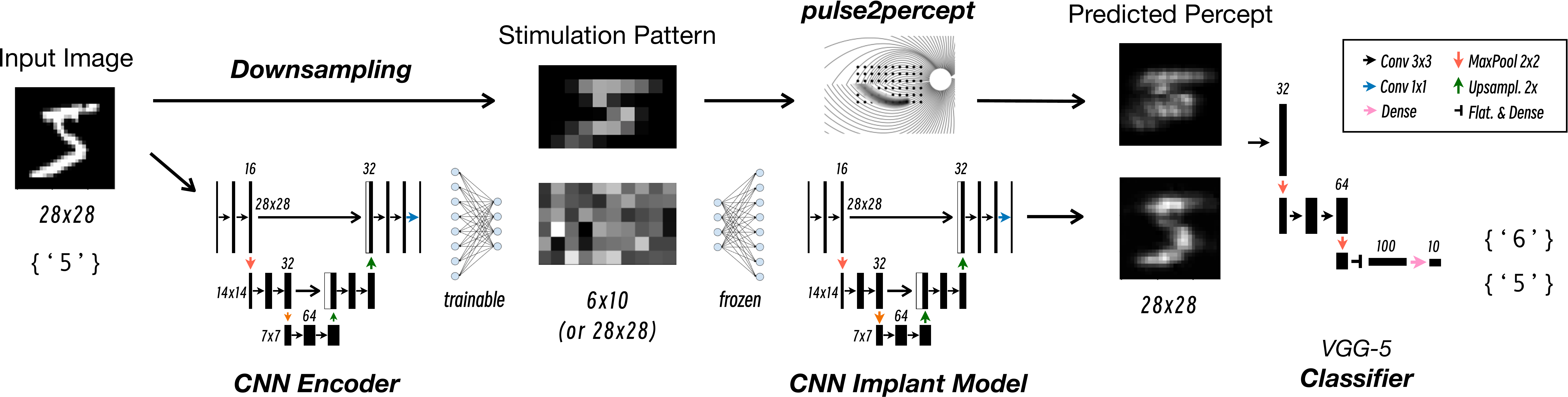}
  \caption{Optimization framework. The upper pipeline illustrates the previous approach using downsampling, where the input image is resized to the desired resolution of the electrode array. The functionality of the retinal implant is simulated by the library \textit{pulse2percept}, which predicts the percept. The lower pipeline illustrates the deep learning-based approach. The input image is encoded with a trainable CNN model, and the corresponding percept is predicted with a pre-trained and frozen CNN implant model. Finally, the percepts are evaluated with a pre-trained VGG-5 classifier.}
  \label{fig:pipeline}
\end{figure*}

\section{MATERIALS AND METHODS}
\subsection{Dataset}\label{sec:dataset}
We used the MNIST dataset~\cite{lecun2010mnist} as input, which contains 70,000 grayscale single handwritten digit images with a resolution of 28$\times$28 and was split into a training set, a validation set and a test set at a ratio of 5:1:1. 
To generate the predicted percepts based on the simulation library \textit{pulse2percept}~\cite{michael_beyeler-proc-scipy-2017}, the \textit{Axon Map} model~\cite{beyeler2019model} was applied with the default parameters. 
The exponential decay constant along the axon $\lambda$ (or axonal decay rate \cite{michael_beyeler-proc-scipy-2017}\cite{granley2022hybrid}) and the exponential decay constant away from the axon $\rho$ (or radial decay rate \cite{michael_beyeler-proc-scipy-2017}\cite{granley2022hybrid}) were trivially chosen as 500 and 150, respectively (cf.~\cite{relic2022deep}\cite{granley2022hybrid}, where different parameters were evaluated). 
Two different resolutions of the stimulation patterns were tested, namely, 28$\times$28 and 6$\times$10. The detailed implementation of the corresponding CNN encoder and CNN implant model is introduced in Sections~\ref{sec:pretrain} and \ref{sec:encoder}.
The predicted percept images were resized to 28$\times$28, which, together with the original images, built a dataset for pre-training a CNN implant model. Both original and percept images were ranged to $[0,1]$.

\subsection{Pre-trained Models}\label{sec:pretrain}
Two models, the U-Net~\cite{ronneberger2015u} implant model and the VGG~\cite{simonyan2014very} classifier, were pre-trained and frozen during the training of the U-Net encoder. 
First, the U-Net implant model mimics the functionality of the \textit{Axon Map} model~\cite{beyeler2019model} implemented in the library \textit{pulse2percept}~\cite{michael_beyeler-proc-scipy-2017} to facilitate an end-to-end deep learning framework. 
We used a shallow U-Net with two down- and upsampling steps. 
On the generated image-to-image dataset introduced in Section~\ref{sec:dataset}, the U-Net implant model was separately pre-trained with a pixelwise mean squared error loss, which converged to $5.93\times10^{-4}$ on the validation set.
In the case of lower resolution stimulation patterns (6$\times$10), an extra fully-connected layer was trained together with the U-Net implant model (Fig.~\ref{fig:pipeline}) and frozen during the optimization of the U-Net encoder. This layer took the downsampled MNIST images as inputs and can be considered to be a trainable upscaler. The lower resolution stimulation patterns were flattened before and after this 60 (i.e. 6$\times$10) to 784 (i.e. 28$\times$28) fully-connected layer.

Second, a VGG-5 classifier was pre-trained with the original MNIST dataset, which evaluates the quality of the encoder with the classification accuracy of the predicted percepts. VGG-5 follows the naming convention in \cite{simonyan2014very} and consists of 3 convolutional layers and 2 fully-connected layers with maxpooling layers in between (Fig.~\ref{fig:pipeline}). Compared to the experiment involving human subjects~\cite{fauvel2022human}, this feedforward VGG classifier not only accelerates the evaluation with quantified metrics, but also directly contributes to the backpropagation through two frozen models to optimize the CNN encoder. Following the recognition approach (see Section~\ref{sec:loss}), two F1-Scores (micro and weighted) were reported to evaluate the performance of the upstream encoder with the two different loss functions mean squared error (MSE) and cross-entropy (CE).

\begin{table}[b]
\centering
\caption{F1-Scores of the MNIST classification. Two different stimulation pattern resolutions (Stim. Res.) were tested with either trivial downsampling or a CNN encoder.}
\label{tab1}
\begin{tabular}{lllrr}
\toprule
Stim. Res. & Encoder & Loss & Micro F1 & Weighted F1  \\ \midrule
$28\times 28$ & None & - &  77.28\%  &      76.63\%   \\ 
$28\times 28$ & U-Net & MSE & 98.08\%  &     98.07\%       \\ 
$28\times 28$ &  U-Net  & CE &  \textbf{98.81\%}  &    \textbf{98.81\%}  \\ \midrule
$6\times 10$ &  Downsampled  & - &  61.87\% &     60.68\%    \\ 
$6\times 10$ &  U-Net  & MSE &  86.37\%  &     85.78\%   \\ 
$6\times 10$  & U-Net  &  CE& \textbf{96.84\%} &     \textbf{96.85\%}    \\
\bottomrule
\end{tabular}
\end{table}

\subsection{CNN Encoder}\label{sec:encoder}
Another U-Net~\cite{ronneberger2015u} serves as the learnable encoder from the input images to the stimulation patterns, which is structurally identical to the U-Net implant model. Analogous to the U-Net implant model, there is an extra fully-connected layer to decrease the output size of the encoder if and only if the desired resolution of the stimulation pattern is lower (6$\times$10 in our experiments). In contrast, this 784 to 60 fully-connected layer is trainable together with the optimization of the U-Net encoder.

\begin{figure}[t]
     \centering
     \begin{subfigure}[b]{0.48\linewidth}
         \centering
         \includegraphics[width=\linewidth]{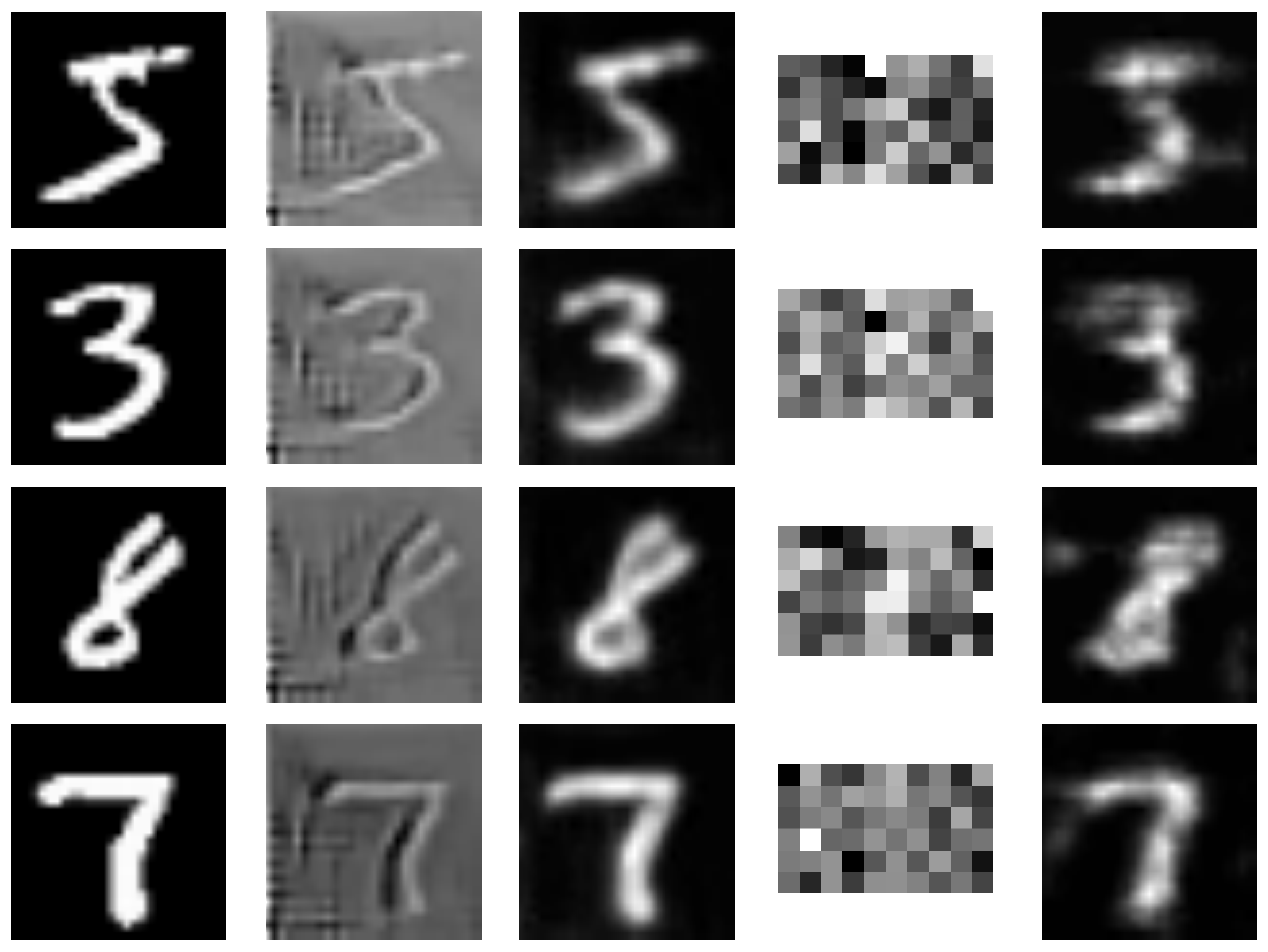}
         \caption{MSE}
     \end{subfigure}
     \hfill
     \begin{subfigure}[b]{0.48\linewidth}
         \centering
         \includegraphics[width=\linewidth]{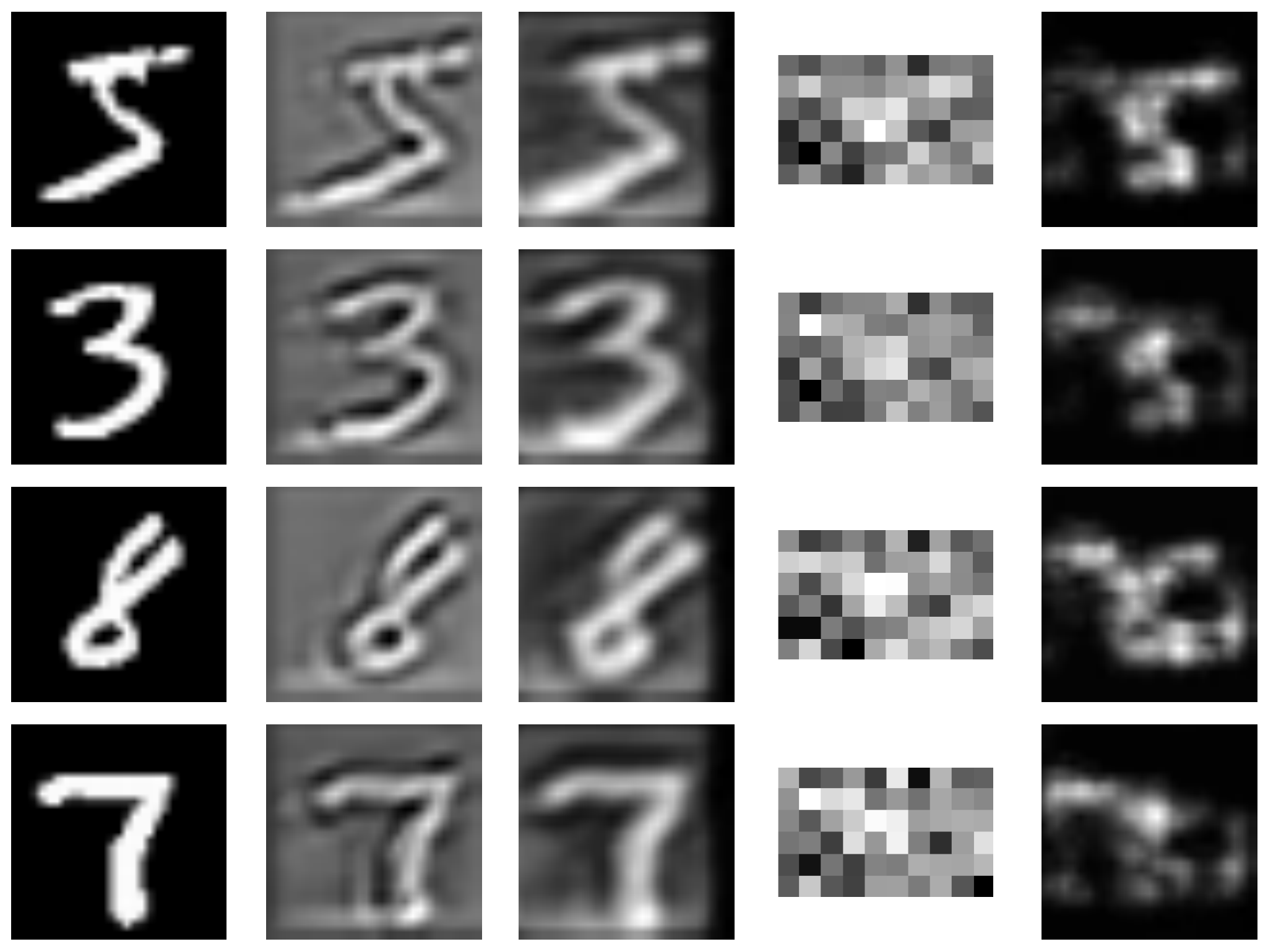}
         \caption{CE}
     \end{subfigure}
        \caption{Visualization with two loss functions: MSE and CE. From left to right: 1. Original images; 2. CNN encoded stimulation patterns of 28$\times$28; 3. Predicted percepts with 28$\times$28 stimuli; 4. CNN encoded stimulation patterns of 6$\times$10; 5. Predicted percepts with 6$\times$10 stimuli.}
        \label{fig:vis}
\end{figure}
\section{EXPERIMENTS AND RESULTS}

\subsection{Higher Accuracy with CNN Encoder}
As shown in TABLE~\ref{tab1}, both high and low resolution stimulation patterns perform considerably better with the U-Net encoder compared to trivial downsampling. Not surprisingly, low resolution stimulation patterns yield less accurate classification results due to the information bottleneck, while the margin between a 784-electrode and a 60-electrode retinal implant narrowed from 15.95\% (no encoder or downsampling) to 1.96\% (CNN encoder with CE) \textit{w.r.t.} the weighted F1-Score. This shows that CNN encoder can extract and transmit information more efficiently at the bottleneck. Furthermore, there is an improvement of 36.17\% \textit{w.r.t.} the weighted F1-Score from downsampling to a CNN encoder on a 60-electrode retinal prosthesis, such as Argus{\textregistered} II. As the MNIST dataset is very balanced, the differences between the micro F1-Scores and the weighted F1-Scores are small. The encoded stimulation patterns of two resolutions and their corresponding predicted percepts are shown in Fig.~\ref{fig:vis}.

\subsection{Reconstruction vs. Recognition}\label{sec:loss}
Previous work tended to use a reconstruction loss to learn and evaluate the predicted percept \cite{van2022end}\cite{relic2022deep}\cite{granley2022hybrid}. In this experiment, we compared two different approaches, namely, \textit{reconstruction} and \textit{recognition}. In the reconstruction approach, an MSE loss is used to minimize the pixelwise difference of the predicted percept and the original input image, while a multi-class CE loss is applied in the recognition approach with a feedforward VGG-5 classifier \cite{simonyan2014very}. 
The recognition experiment is based on the assumption that, even though the phosphene percept is not perfectly identical to the original image, the information can still be recognized by the visual cortex, as long as the semantic label can be classified by a shallow VGG.
Thus, inspired by Lindsey and Ocko et al.~\cite{lindsey2018unified}, the recognition approach can effectively be considered as loosened processing, which shifts learnability from the retina or retinal implant (U-Net implant model) to the visual cortex (VGG-5 classifier).

The approach using a CE loss outperforms the MSE in the weighted F1-Score by 0.74\% with the 28$\times$28 and by 11.07\% with the 6$\times$10 stimulation pattern, respectively (TABLE~\ref{tab1}).
However, the visual quality between the CE and the MSE loss does not in correspond to the quantitative results, as shown in Fig.~\ref{fig:vis}. 
The loss function plays an important role in where the training processing places its focus: MSE tries to minimize the pixelwise difference, which is associated with the visual semantics, and CE tries to minimize a more abstract and higher-level criterion, namely, the classification.

\subsection{Biomimicry of CNN Encoder}
 Ideally, the epiretinal implant~\cite{walter2005epiretinal} should contain the functionality of the retinal ganglion cells. To the best of our knowledge, the implant model implemented in \textit{pulse2percept}~\cite{michael_beyeler-proc-scipy-2017} does not involve the specific functionality of the retinal ganglion cells (RGC).
 We found that the encoded stimulation patterns with a U-Net are visually more similar to the Difference of Gaussians (DoG), which is a computational model of the RGC layer \cite{rodieck1965quantitative}\cite{hartline1969visual}, than the original image. Therefore, we calculated two metrics~\cite{hore2010image} regarding the similarity of images, namely, peak signal-to-noise ratio (PSNR) and structural similarity index measure (SSIM)~\cite{wang2004image}.
 To simplify the computation, we computed the Laplacian with the following filter
\begin{center}
$\begin{bmatrix}
0 & 1 & 0\\
1 & -4 & 1\\
0 & 1 & 0 \\
\end{bmatrix}$,
\end{center}
which approximates a specific example of DoG. As listed in TABLE~\ref{tab2}, the CNN encoded stimulation patterns are more similar to the Laplacian-filtered images than the original images \textit{w.r.t.} PSNR and SSIM. This shows that the CNN encoder can be learned to behave biomimetically without any specific constraint.

\begin{table}[b]
\centering
    \caption{PSNR and SSIM of the CNN encoded stimulation patterns to the original or to the biomimetic (approximated by a Laplacian filter) images with a resolution of 28$\times$28. The two metrics between the original and the Laplacian-filtered images are reported as a baseline.}
    \label{tab2}
\begin{tabular}{llll}
\toprule
 Loss & Comparate & PSNR ($\uparrow$)  &  SSIM ($\uparrow$) \\ \midrule
MSE & Original &   55.186  &    +0.027    \\ 
MSE & Laplacian &  \textbf{62.822}    &   \textbf{+0.179}   \\\midrule
CE & Original &  54.840    &  -0.069    \\ 
CE & Laplacian &    \textbf{61.872}  &   \textbf{+0.284}   \\\midrule
\multicolumn{2}{l}{cf. Original-Laplacian} &53.307& -0.230\\
\bottomrule
\end{tabular}
\end{table}

\subsection{Correlation of Low Resolution Stimuli}
The encoded stimulation patterns are not always human-understandable, especially when the resolution of the retinal prosthesis is limited (as visualized in Fig.~\ref{fig:vis} in the fourth column). While trying to evaluate the correlation of the stimuli of low resolution, we calculated the cosine similarity $\mathrm{S_c}(s_1, s_2)= ({{s_1}^T\cdot s_2})/({ \|s_1\|\| s_2\|})$, with $s_1$ and $s_2$ being the pairwise normalized and flattened stimulation patterns.

\begin{figure}[t]
     \centering
     \begin{subfigure}[b]{0.47\linewidth}
         \centering
         \includegraphics[width=\linewidth]{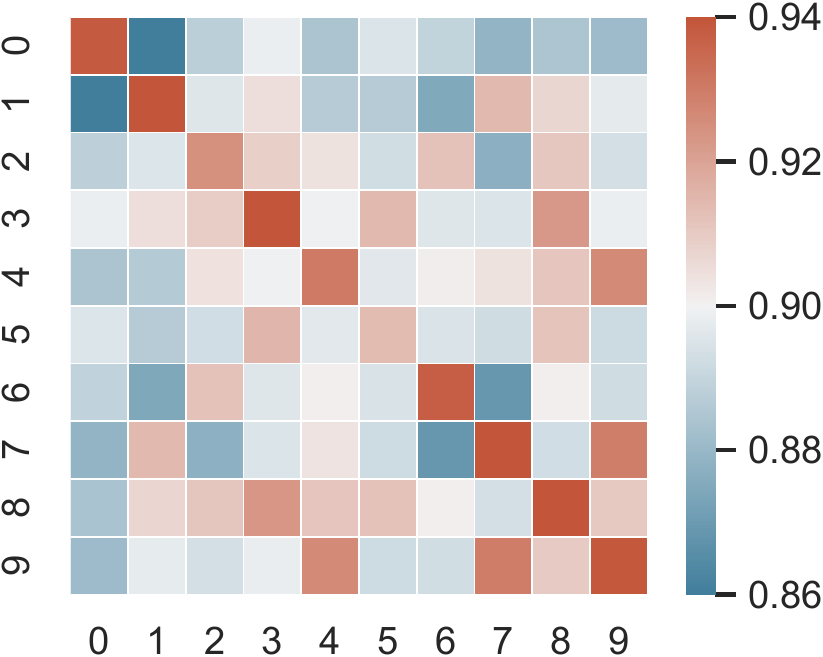}
         \caption{MSE}
         \label{fig:y equals x}
     \end{subfigure}
     \hfill
     \begin{subfigure}[b]{0.47\linewidth}
         \centering
         \includegraphics[width=\linewidth]{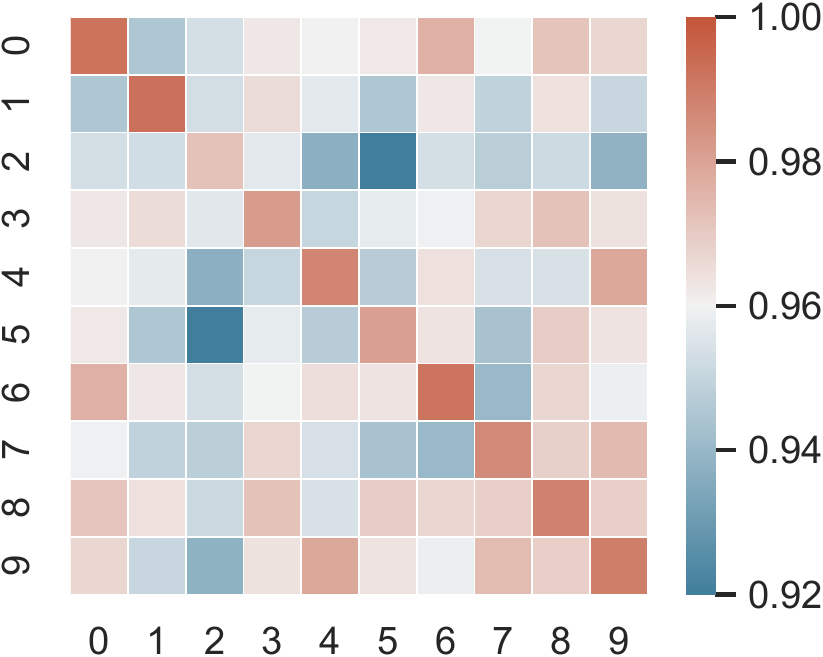}
         \caption{CE}
         \label{fig:three sin x}
     \end{subfigure}
        \caption{Cosine similarity between the normalized and flattened stimulation patterns of 6$\times$10 in two confusion matrices of MSE and CE, respectively. The colors are based on the cosine similarity range in each approach.} 
        \label{fig:confusion}
\end{figure}

As illustrated in Fig.~\ref{fig:confusion}, the stimuli belonging to the same class show a higher cosine similarity.
Moreover, the cosine similarity between the digits 0 and 1 is the lowest in the approach trained with MSE, and between the digits 2 and 5 in the approach trained with CE, which is in correspondence with the handwritten trajectory of high and low intensities.

\section{DISCUSSION}

We have shown that the CNN encoder performs by far better than the trivial downsampling approach. We believe that deep learning techniques, due to their accuracy and generalizability, will play an important role in the next generation of visual prostheses. 

As mentioned in Section~\ref{sec:dataset}, the default pair of parameters in \cite{michael_beyeler-proc-scipy-2017} ($\lambda=500$ and $\rho=150$) is the only one used to generate the dataset for training the CNN implant model, while, for example, \cite{relic2022deep} and \cite{granley2022hybrid} utilize multiple sets of patient-specific parameters to train the surrogate implant model. We expect to verify more real-world parameters and more computational models \cite{beyeler2019model}\cite{granley2021computational}\cite{van2022biologically} with the fully neural network-based framework. In addition, more complicated datasets can be involved, where the instance-level bounding box is extracted beforehand and the region of interest (RoI) is fed into the proposed framework.

\addtolength{\textheight}{-12cm}   




\bibliographystyle{IEEEtran} 
\bibliography{IEEEabrv,bib}

\end{document}